\documentclass[prl,amsmath,showpacs,superscriptaddress,amssymb,twocolumn,floatfix,aps]{revtex4}
\usepackage{amssymb}
\usepackage{epsfig}
\usepackage{ulem} 
\usepackage[usenames]{color}

\newcommand{\be}{\begin{eqnarray}}
\newcommand{\ee}{\end{eqnarray}}

\def \figwidth{8cm}

\begin{document}

\title{Model Independent Extraction of S-Matrix Poles from Experimental Data}

\author{S. Ceci\footnote{Email: sasa.ceci@irb.hr}}
\affiliation{Rudjer Bo\v{s}kovi\'{c} Institute, Bijeni\v{c}ka  54, HR-10000 Zagreb, Croatia}


\author{M.~Korolija}
\affiliation{Rudjer Bo\v{s}kovi\'{c} Institute, Bijeni\v{c}ka  54, HR-10000 Zagreb, Croatia}


\author{B.~Zauner}
\affiliation{Rudjer Bo\v{s}kovi\'{c} Institute, Bijeni\v{c}ka  54, HR-10000 Zagreb, Croatia}

\date{\today}


\begin{abstract}
By separating data points close to a resonance into intervals, and fitting all possible intervals to a simple pole with constant coherently added background, we obtained a substantial number of convergent fits. After a carefully chosen set of statistical constraints was imposed, we calculated the average of a resonance pole position from the statistically acceptable results. We used this method to find pole positions of $Z$ and N(1440) resonances, and to show that the strong discrepancy between the old and new measurements of the $\Upsilon$(11020) mass stems from specious comparison of the $\Upsilon$(11020) pole with its Breit-Wigner mass.
\end{abstract}
  

\maketitle



Breit-Wigner (BW) parameters are often used for the description of unstable particles (see e.g.~{\it Review of Particle Physics} \cite{PDG08}), although shortcomings of such choice have been pointed out on numerous occasions. For example, Sirlin showed that the BW parameters of the $Z$ boson were gauge dependent \cite{Sir91}. To resolve this issue he redefined BW parameters, but also suggested usage of the S-matrix poles as an alternative, since poles are fundamental properties of the S-matrix and therefore gauge independent by definition. In a somewhat different study, H\"ohler advocated using S-matrix poles for characterization of nucleon resonances \cite{Hoh93} in order to reduce confusion that arises when different definitions of BW parameters are used \cite{NstarPWAs}. However, loosely defined \cite{Hoh00} BW parameters of mesons and baryons are still being extracted from experimental analyses, compared among themselves \cite{PDG08}, and used as input to QCD-inspired quark models \cite{QM} and as experiment-to-theory matching points for lattice QCD \cite{Dur08}. 

The main motivation for this research are strong discrepancies between the old and newly obtained parameters of some well known resonances. In particular, {\sc BaBar} collaboration recently reported that the mass of $\Upsilon(11020)$ is 10996$\pm$2~MeV \cite{Aub09}, significantly different from the old value of 11019$\pm$2~MeV \cite{PDG08}. Furthermore, the width turned out to be less than a half of the expected, namely 37$\pm$3~MeV instead of 79$\pm$16~MeV. 

In this paper, we developed a reliable method for model-independent extraction of S-matrix pole positions directly from the data, and connected them to the Breit-Wigner parameters in order to understand the observed discrepancies. We showed that the both results, {\sc BaBar} and PDG, are consistent: the old ones should be interpreted as BW mass and width, while the new ones are pole parameters.


The first step in devising a method for extraction of the pole parameters from the experimental data is to set up an appropriate parameterization. The parameterization presented here is based on the assumption that close to a resonance, the T matrix will be well described with a simple pole and a constant background. The similar assumption was used in H\"ohler's speed plot technique \cite{Hoh93}. The speed plot is a method used for the pole parameter extraction from the known scattering amplitudes. It is based on calculating the first order energy derivative of the scattering amplitude, with the key assumption that the first derivative of the background is negligible. 

The T matrix with a single pole and constant background term is given by
\begin{equation}
T(W) =r_p\,\frac{\Gamma_p/2}{M_p-W-i\,\Gamma_p/2}+b_p,
\label{eq:parameterizationA}
\end{equation}
where $W$ is center-of-mass energy, $r_b$ and $b_b$ are complex, while $M_p$ and $\Gamma_p$ are real numbers. Total cross section is then given by $\sigma\approx|T|^2/q^2$, where $q$ is the initial center-of-mass momentum. Equation (\ref{eq:parameterizationA}), as well as other similar forms (see e.g.~\cite{PDG08}), are standardly called Breit-Wigner parameterizations, which can be somewhat misleading since $M_p$ and $\Gamma_p$ are generally not Breit-Wigner, but pole parameters (hence the index $p$). The square of the T matrix defined in Eq.~(\ref{eq:parameterizationA}) is given by
\begin{equation}\label{eq:parameterization}
|T(W)|^2=T_\infty^2\,
\frac
{
(W-M_z)^2+\Gamma_z^2/4
}
{
(W-M_p)^2+\Gamma_p^2/4
},
\end{equation}
where, for convenience, we simplified the numerator by combining the old parameters into three new real-valued ones: $T_\infty$, $M_z$, and $\Gamma_z$. Pole parameters $M_p$ and $\Gamma_p$ are retained in the denominator.

With such a simple parameterization, it is crucial to use only data points close to the resonance peak. To avoid picking and choosing the appropriate data points by ourselves, we analyzed the data from a wider range around the resonance peak, and fitted localy the parameterization (\ref{eq:parameterization}) to each set of seven successive data points (seven data points is minimum for our five-parameter fit). Then we increased the number of data points in the sets to eight and fitted again. We continued increasing the number of data points in sets until we fitted the whole chosen range. Such procedure allowed different background term for each fit, which is much closer to reality than assuming a single constant background term for the whole chosen data set (see e.g.~discussion on the problems with speed plot in Ref.~\cite{Ceci08}). In the end, we imposed a series of statistical constraints to all fits to distinguish the good ones.

In order to pinpoint the statistical strategy to be used, we did a substantial number of simulations with the data sets that had known poles and zeros. It turned out that the most successful strategy was to make an ordered list of all fit results, from best to worst, and then to drop the worst three quarters using the following goodness-of-fit measures: Akaike information criterion \cite{AIC}, Schwartz (Bayesian information) criterion \cite{BIC}, and P-values of the extracted fit parameters (in particular, $M_p$ and $\Gamma_p$). Eventually, we kept the intersection of the fits that satisfied all criteria. Results closest to the original poles were produced by averaging the obtained pole positions of all good fits. The standard deviation turned out to be a good estimate for errors of obtained parameters. All other approaches we tested, such as keeping only a handful of the best fits, or keeping just those whose values of reduced $\chi^2$ were close to one, failed to accurately reproduce the originalpole parameters. The whole analysis was done in Wolfram~Mathematica~7 using NonlinearModelFit routine \cite{wolfram}.

\begin{figure}[h!]
\begin{center}
\includegraphics[width=\figwidth]{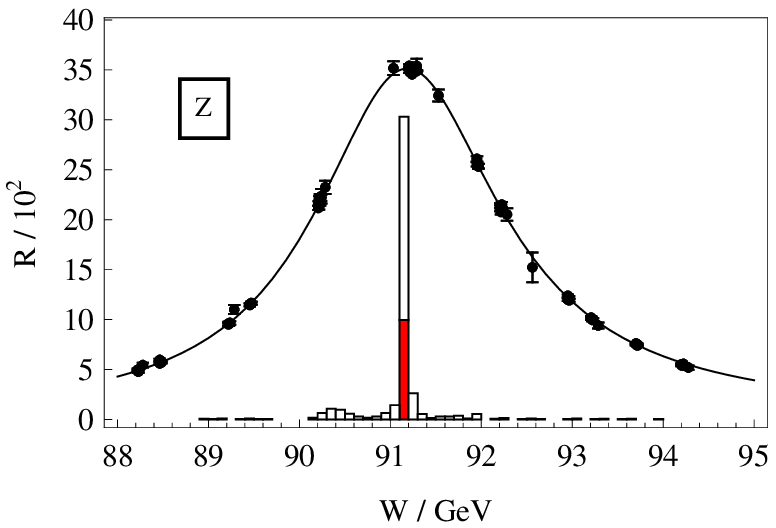}
\includegraphics[width=\figwidth]{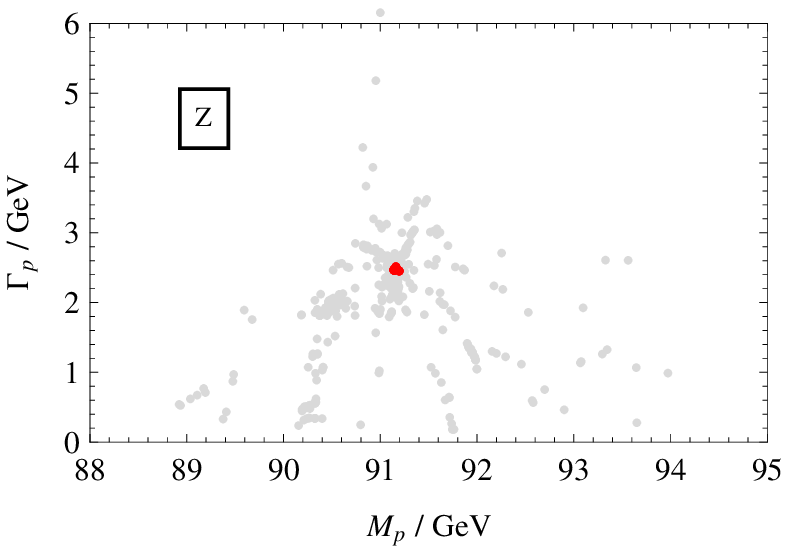}
\caption{ [Upper figure] PDG compilation of $Z$ data \cite{PDG08} and histogram of obtained pole masses. Line is the fit result with the lowest reduced $\chi^2$ (just for illustration). Dark (red online) colored histogram bins are filled with statistically preferred results. [Lower figure] Pole masses vs. pole widths. Dark (red online) circles show statistically preferred results we use for averages.}
\label{fig:Z0fig1}
\end{center}
\end{figure}

Having defined the fitting strategy, we tested the method by applying it to the case of the $Z$ boson. The data set we used is from the PDG compilation \cite{PDG08}, and shown in Fig.~\ref{fig:Z0fig1}. Extracted pole masses are shown in the same figure: filled histogram bins show pole masses from the good fits, while the empty histogram bins are stacked to the solid ones to show masses obtained in the discarded fits. Height of the pole-mass histogram in Fig.~\ref{fig:Z0fig1} is scaled for convenience.

Extracted S-matrix pole mass and width of $Z$ boson are given in Table~\ref{tbl:table1}. The pole masses are in excelent agreement, while the pole widths are reasonably close. It is important to stress that the difference between the pole and BW mass of the Z boson is fundamental and statistically significant. Distribution of discarded and good results is shown in the lower part of Fig.~\ref{fig:Z0fig1}.

\begin{table}[h!]
\caption{ Pole parameters of $Z$ obtained in this work. PDG values of pole and BW parameters are given for comparison.}
\label{tbl:table1}
\begin{ruledtabular}
\begin{tabular}{lrrr}
\fbox{$Z$}  & Pole & Pole PDG \cite{PDG08}  & BW PDG \cite{PDG08}\\
\cline{2-2}\cline{3-3}\cline{4-4}
$M$/MeV       &  91159 $\pm$ 8    & 91162 $\pm$ 2    & 91188 $\pm$ 2\\ 
$\Gamma$/MeV  &   2484 $\pm$ 10   &  2494 $\pm$ 2    & 2495 $\pm$ 2\\
\end{tabular}
\end{ruledtabular}
\end{table}

Next, we turn to the data from {\sc BaBar} collaboration \cite{Aub09} to determine whether the PDG averages for $\Upsilon$(11020), or the newly reported resonance parameters obtained in Ref.~\cite{Aub09} are correct.

\begin{figure}[h]
\begin{center}
\includegraphics[width=\figwidth]{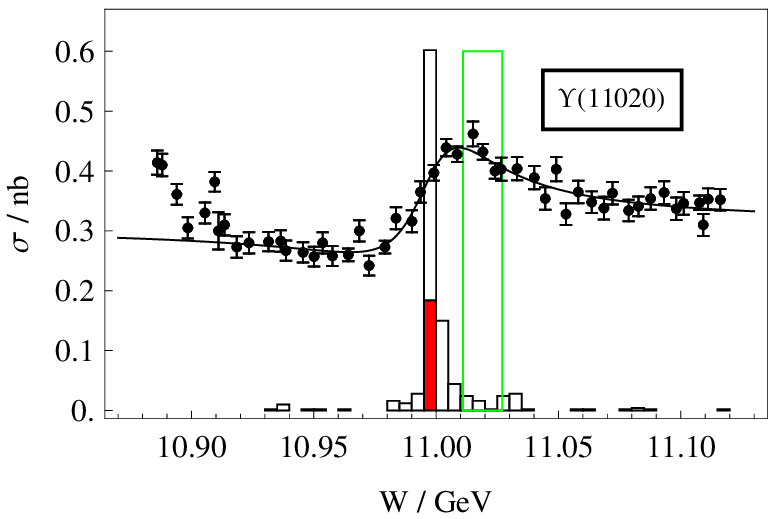}
\includegraphics[width=\figwidth]{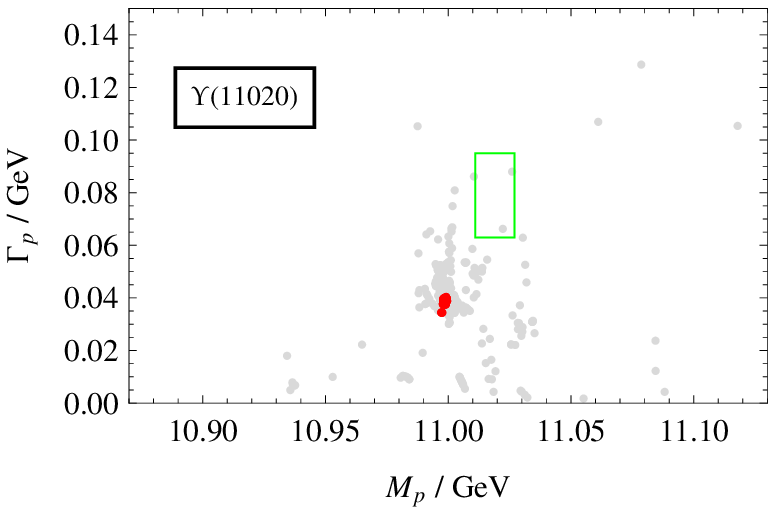}
%
\caption{$\Upsilon$(11020) resonance pole obtained by our method. Grey (green online) rectangles represents PDG range for $\Upsilon$(11020) mass (both figures) and width (only lower figure).}
\label{fig:Y11020fig2}
\end{center}
\end{figure}

Our local fit of the $\Upsilon$(11020) pole parameters is shown 
in Fig.~\ref{fig:Y11020fig2}. As in the case of $Z$ boson, 
the full and empty histograms show how many of the extracted pole mass fits 
were accepted or discarded in the analysis. 

Average values obtained for the resonance mass and the 
width are given in Table~\ref{tbl:table2b},
together with the same parameters obtained in the {\sc BaBar} analysis, and those quoted by PDG. Extracted pole parameters of $\Upsilon$(11020) are practically the same as those reported in \cite{Aub09}, even though our parameterization is much simpler (single pole plus constant background vs.~two constant background, and two pole terms). The original results cited in PDG (from CUSB \cite{CUSB}, and CLEO \cite{CLEO}) were obtained by fitting Gaussians to the resonance peaks in the data, and peak positions are usually closer to the BW mass.

\begin{table}[h]
\caption{Parameters of $\Upsilon$(11020) meson. Pole parameters are results of this work.}
\label{tbl:table2b}
\begin{ruledtabular}
\begin{tabular}{lrrr}
\fbox{$\Upsilon$(11020)}  & Pole & {\sc BaBar} \cite{PDG08,Aub09}  & PDG \cite{PDG08} \\
\cline{2-2}\cline{3-3}\cline{4-4}
$M$/MeV      &  10999 $\pm$ 1   & 10996 $\pm$ 2    & 11019 $\pm$ 8\\ 
$\Gamma$/MeV &     38 $\pm$ 1   &    37 $\pm$ 3    &     79 $\pm$ 16\\
\end{tabular}
\end{ruledtabular}
\end{table}


\begin{table}[h]
\caption{N(1440) resonance parameters.}
\label{tbl:table3}
\begin{ruledtabular}
\begin{tabular}{lrrr}
{\fbox{$N$(1440)}}  & Pole  & Pole PDG \cite{PDG08}  & BW PDG \cite{PDG08} \\
\cline{2-2}\cline{3-3}\cline{4-4}
$M$/MeV      & 1370  $\pm$ 6  & 1365 $\pm$ 15  & 1440 $\pm$ $^{30}_{20}$\\ 
$\Gamma$/MeV &  197  $\pm$ 6  &  190 $\pm$ 30  &  300 $\pm$ $^{150}_{100}$\\
\end{tabular}
\end{ruledtabular}
\end{table}

To investigate this case further, we analyze another resonance with strong difference between the pole and BW mass, the Roper resonance N(1440). We extracted Roper resonance pole parameters from the $\pi N$ elastic $P_{11}$ partial wave obtained in the GWU analysis \cite{Arn10}. According to PDG, this wave has a very rich structure: there is a four-star Roper resonance, a three-star N(1710) resonance, and a one-star N(2100) resonance. However, the GWU analysis reports only one resonance in this partial wave, the Roper N(1440) resonance. In a preliminary analysis, we could see some indication for all resonances mentioned by PDG but, for this study, we focus on N(1440) because of its unusually strong shift between the pole and BW mass (roughly 75~MeV).

\begin{figure}[h]
\begin{center}
\includegraphics[width=\figwidth]{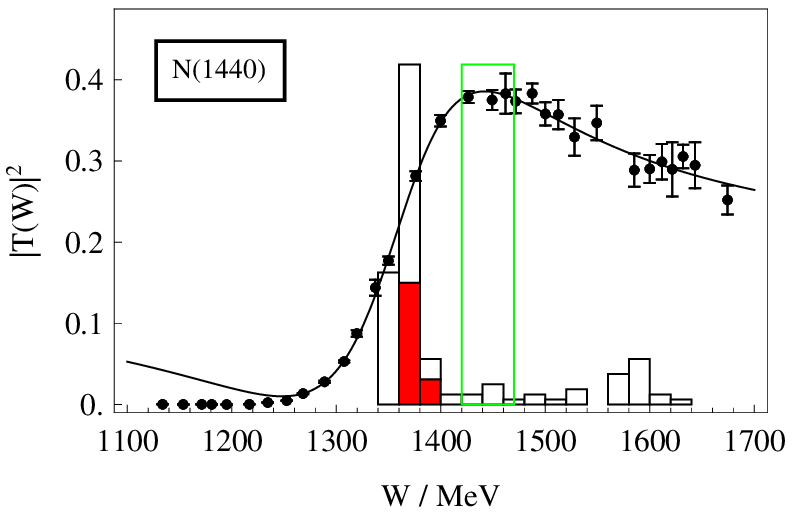}
\includegraphics[width=\figwidth]{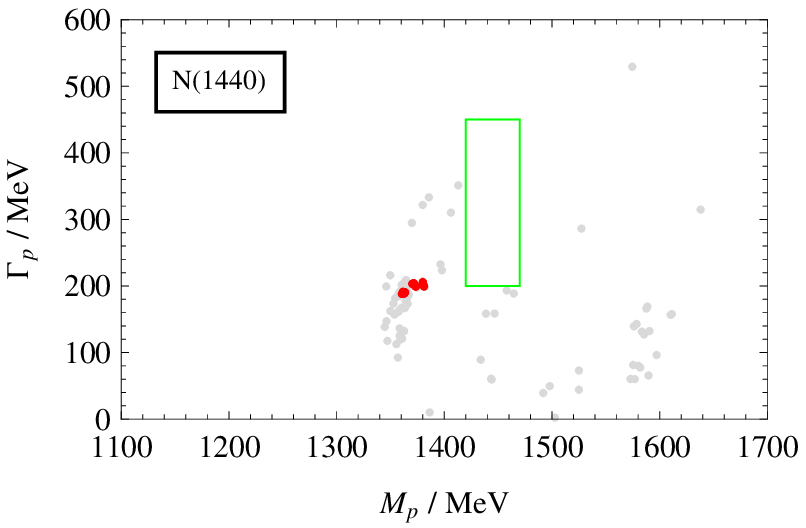}
\caption{N(1440) resonance pole obtained by our method. Grey (green online) rectangles represents PDG range for N(1440) Breit-Wigner mass (both figures) and width (only lower figure).}
\label{fig:N1440fig1}
\end{center}
\end{figure}

Our results for N(1440) are given in Table \ref{tbl:table3}, where we see that the pole parameters are in an excellent agreement with the PDG estimates. Unlike the pole mass, BW masses are situated closer to the positions of the peak (see Figs.~\ref{fig:Y11020fig2} and \ref{fig:N1440fig1}). 

%

The field-theory reason for the resonance pole shift is the energy dependence of the imaginary part of resonance self energy, which is commonly modeled by the energy dependent width \cite{Flatte,Man95}. Equation (\ref{eq:parameterizationA}) would be exact if the self energy was constant. In more realistic cases, the T-matrix denominator $D(W)$ is given by 
\begin{equation}
D(W)=M_b-W-i\,\Gamma(W)/2, 
\end{equation}
where we introduce the BW mass and $M_b$, which is generally not the real part of the pole position. Keeping only the first two terms in Taylor expansion of $\Gamma(W)$ about $W=M_b$ (as done in Ref.~\cite{Man95}) the width becomes
\begin{align}
\Gamma(W)/2 &= \Gamma_b/2+\tan\theta\,\,(W-M_b),
\end{align}
where the $\Gamma_b$ is a shorthand for $\Gamma(M_b)$, and $\tan\theta$ is the slope of $\Gamma/2$ at $W=M_b$. The pole position $M_p~-~i\Gamma_p/2$ is obtained by solving $D(W)=0$, which yields
\begin{align}
 M_p &= M_b+\sin\theta\,\cos\theta\,\Gamma_b/2,\label{eq:Mp}\\
 \Gamma_p &=\cos^2\theta\,\Gamma_b.\label{eq:gammap}
\end{align}

%

Relations (\ref{eq:Mp}) and (\ref{eq:gammap}), originally introduced in Ref.~\cite{Man95}, may be used to cross check estimates for pole and BW parameters. Table \ref{tbl:table4} shows angles $\theta$ for all resonances analyzed in this paper, calculated from PDG estimates for pole and BW widths using Eq.~(\ref{eq:gammap}). Since $M_p$ is smaller than $M_b$, we chose negative $\theta$ solution (cf.~Eq.~(\ref{eq:Mp}). It turns out that the {\sc BaBar} value of $\Upsilon$(11020) mass is accurately reproduced.

\begin{table}[h]
\caption{The connection between S-matrix pole and Breit-Wigner parameters using only the PDG values.   }
\label{tbl:table4}
\begin{ruledtabular}
\begin{tabular}{lrrr}
  & $\theta$/$^\circ$  & $M_p$/MeV PDG\cite{PDG08} & $M_p$/MeV Eq.~(\ref{eq:Mp}) \\
\cline{2-2}\cline{3-3}\cline{4-4}    
$\Delta$(1232)    & -23.0    &  1210 $\pm$ 1                               & 1210\\
$N$(1440)         & -37.3    &  1365 $\pm$ 15                              & 1368\\
$\Upsilon$(11020) & -46.8    &  10996\footnote{{\sc BaBar} value.} $\pm$ 2 & 10999\\
$Z$               & -1.26    &  91162 $\pm$ 2                              & 91161

\end{tabular}
\end{ruledtabular}
\end{table}
Does this $\theta$ carry any physical meaning? For resonances with one dominant decay channel, such as the $\Delta$(1232), we can impose a unitarity condition ($\mathrm{Im}\,T~=~|T|^2$) to Eq.~(\ref{eq:parameterizationA}) and learn that $r_p=e^{2i\theta}$, and $b_p=e^{i\theta}\sin\theta$. It is the same $\theta$ and represents a half of the complex residue phase. Indeed, from Ref.~\cite{PDG08} we read that $\Delta$(1232) has $(-47\pm1)^\circ$ for pole residue phase, quite consistent with -46$^\circ$, a double value of the $\theta$ from Table \ref{tbl:table4}. However, this simple relation is lost when important inelastic channels are open, e.g.~in the N(1440) case, where 2$\,\theta\approx$~-75$^\circ$, which is significantly larger than its residue phase -100$^\circ$ \cite{PDG08}. The difference comes from different $\Gamma(W)$ in the denominator and numerator of T matrix: total decay width is in the denominator, while the partial widths are in the numerator. Since the energy dependence of the two is in general different, their slopes (i.e.~$\tan\theta$) will be different as well.  

Since our pole extraction method confirmed {\sc BaBar} result, the successful cross check is the last piece of the puzzle. PDG estimates of $\Upsilon$(11020) are consistent with BW parameters.

%
In conclusion, we have developed a model-independent method for extraction of 
resonance pole parameters from total cross sections and partial waves. Very good estimates for $Z$ boson, $\Upsilon$(11020), and N(1440) pole positions were obtained. Furthermore, we showed that the strong discrepancy between PDG estimates and {\sc BaBar} result for $\Upsilon$(11020) comes from specious comparison of the pole and BW mass. 

We are today witnessing the dawn of ab-initio calculations in low-energy QCD. In order to compare theoretical predictions with experimentally determined resonance states, we need first to establish proper point of comparison. The case of $\Upsilon$(11020) is a vivid example how particularly careful we must be when choosing this point. Therefore, we would like to express our concern about other potentially problematic comparisons between the pole and BW parameters in the literature, in particular in the {\it Review of Particle Physics}, and recommend drawing a clear distinction between the two in future publications.

S.C. owes a debt of gratitude to A.~\v Svarc, S.~Krewald, M.~Manley, H.~Haberzettl, M.~D\"oring, A.~Sibirtsev, V.~Brigljevi\'c, S.~Szilner, and N.~Tepi\'c for their support and will to participate in discussions about the topics addressed herein. This work is supported in part by the DAAD (Deutscher Akademischer Austauschdienst) grant No.~D/08/00215 and the DFG (Deutsche For\-schungs\-ge\-mein\-schaft, Gz.:~DO~1302/1-2).

%

\end{document}